\documentclass[twocolumn,preprintnumbers,amsmath,amssymb]{revtex4}
\usepackage{graphicx}
\usepackage{dcolumn}
\usepackage{bm}
\usepackage{hyperref} 
\usepackage{multirow} 
\usepackage{soul}
\newcommand\beq{\begin{equation}}
\newcommand\eeq{\end{equation}}
\newcommand\bea{\begin{eqnarray}}
\newcommand\eea{\end{eqnarray}}
\usepackage{wrapfig}
\newcommand{\nonum}{\nonumber}

\begin{document}

\title{\bf Spin parity and broken symmetry in finite spin-1/2 chains with frustrated exchange: quantum transition from high to low spin \\}
\author{\bf Manoranjan Kumar and Zolt\'an G. Soos }
\address{ Department of Chemistry, Princeton University, Princeton NJ 08544 \\}
\date{\today}

\begin{abstract}
 Exact diagonalization of finite spin-1/2 chains with periodic boundary 
conditions is applied to the ground state (gs) of chains with ferromagnetic (F) exchange $J_1 < 0$ 
between first neighbors, antiferromagnetic (AF) exchange $J_2 = \alpha J_1 > 0$ 
between second neighbors, and axial anisotropy $0 \le \Delta \le 1$. In zero field, 
the gs is in the $S_z = 0$ sector for the relevant parameters and is doubly 
degenerate at multiple points $\gamma_m = (\alpha_m, \Delta_m)$ in the $\alpha$, $\Delta$ plane. 
Degeneracy under inversion at sites or spin parity or both leads, respectively, to a bond order 
wave (BOW), to staggered magnetization or to vector chiral (VC) order. Exact results up to $N = 28$ 
spins directly yield order parameters and spin correlation functions whose weak N dependencies 
allow inferences about infinite chains. The high-spin gs at $J_2 = 0$ changes discontinuously at $\gamma_1 = (-1/4, 1)$ 
to a singlet in the isotropic ($\Delta = 1$) chain. The transition from high to low spin $S(\alpha, \Delta)$ 
is continuous for $ \Delta < \Delta_B = 0.95 \pm 0.01$ on the degeneracy line $\alpha_1(\Delta)$. The gs has 
staggered magnetization between $\Delta_A = 0.72$ and $\Delta_B$, and a BOW for $\Delta < \Delta_A$. When both 
inversion and spin parity are reversed at $\gamma_m$, the correlation functions $C(p)$ for spins 
separated by $p$ sites are identical. $C(p)$ minima are shifted by $\pi/2$ from the minima of 
VC order parameters at separation $p$, consistent with right and left-handed helices along 
the z axis and spins in the xy plane. Degenerate gs of finite chains are related to 
quantum phase diagrams of extended $\alpha$, $\Delta$ chains, with good agreement for order 
parameters along the line $\alpha_1(\Delta)$. Degenerate gs limit a VC phase to intermediate $\alpha$ and $\Delta$ 
where $S(\alpha, \Delta)$ varies rapidly but continuously, in contrast to many-body 
treatments in which VC phases extend over a larger range of parameters.
 
\vskip .4 true cm
\noindent PACS numbers:
75.10.Jm,75.10.Pq, 5.40.Cx,05.30.Rt \\
\noindent Email: soos@princeton.edu,manoranj@princeton.edu
\end{abstract}

\maketitle

\section{Introduction}
One-dimensional (1D) spin chains have been extensively studied 
over the years both experimentally and theoretically. Spin chains are good 
approximations to the magnetism of diverse inorganic and organic crystals. 
They are simple many-body quantum systems, well suited for computational 
studies \cite{r1}, with some exactly known properties and rich 
ground-state (gs) phase diagrams. Solid-state studies focus broadly on 
magnetic properties, instabilities and phase transitions that limit 1D behavior 
at low temperature. Theoretical interest extends to quantum phases with different 
gs in parameter space. In this paper we consider the gs properties of spin-1/2 chains, 
Eq. \ref{eq1} below, with isotropic or axially anisotropic exchange $J_1$ between nearest 
neighbors and antiferromagnetic (AF) exchange $J_2 > 0$ between second neighbors that 
ensures frustration for either sign of $J_1$. The spin-Peierls system \cite{r2} $\rm CuGeO_3$ 
illustrates spin-1/2 chains of Cu(II) ions with $J_1 > 0$. The isotropic AF/AF chain 
is a spin liquid up to \cite{r3} $\alpha = J_2/J_1 \le 0.2411$ and its exact gs is a 
bond order wave (BOW) at $\alpha = 1/2$, the Majumdar-Ghosh point \cite{r4}. \\

\begin{figure}[h]
\begin {center}
\hspace*{-0cm}{\includegraphics[width=7.5cm,height=2.5cm,angle=0]{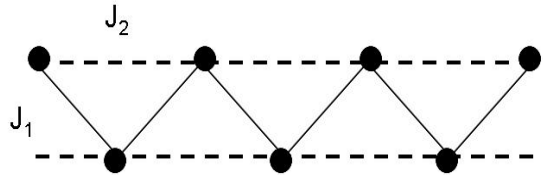}} \\
\caption{Schematic representation of a spin-1/2 chain with exchange $J_1$ and $J_2$ between first and second neighbors. } 
\label{fig1}
\end {center}
\end{figure}

Spin-1/2 Cu(II) chains with ferromagnetic (F) $J_1 < 0$ have recently been identified 
in cupric oxides \cite{r5,r6,r7,r8,r9,r10,r11,r12,r13} with estimated $\alpha$'s ranging 
from \cite{r13} $\alpha \approx 0$ in $\rm Ba_3Cu_3In_4O_{12}$ or $ \rm Ba_3Cu_3Sc_4O_{12}$ 
to \cite{r12,r14} $\alpha \approx -0.5$ in $\rm LiCuSbO_4$, $\rm LiCuZrO_4$, $\rm LiCuVO_4$ 
and $ \rm LiCu_2O_2$. The isotropic F/AF chain has a ferromagnetic gs for $\alpha = J_2/J_1 \ge \alpha_c = -1/4$, 
a singlet gs for $\alpha \le \alpha_c$, and exact degeneracy at the quantum critical point $\alpha_c$ 
as shown by Hamada et al.\cite{r15} Vector chiral (VC)  and multipolar phases have been of special 
interest \cite{r14,r16,r17,r18,r19,r20}. Hikihara et al. \cite{r19} discuss the phase 
diagram of the isotropic F/AF chain in a static magnetic field. Furukawa et al. \cite{r17} 
and Sirker \cite{r14}, among others \cite{r14r}, have studied the axially anisotropic F/AF chain 
in zero field, the model considered in this paper. The limit $J_1 = 0$ 
decouples the system into two AF chains as sketched in Fig. \ref{fig1} for a zigzag chain.\\ 
 
The Hamiltonian of the anisotropic F/AF chain with periodic boundary conditions (PBC) and spin-1/2 sites is
\begin{eqnarray}
H(\alpha,\Delta) & =& J_1 \sum^N_{p=1} (\vec{S}_p \cdot \vec{S}_{p+1} + \alpha \vec{S}_p \cdot \vec{S}_{p+2})  \nonum \\
&+&(\Delta-1)(S^z_p S^z_{p+1} + \alpha S^z_p S^z_{p+2})
\label{eq1}
\end{eqnarray}
\noindent $J_1 = -1$ sets the energy scale. The model has two parameters, the frustration 
ratio $\alpha = J_2/J_1< 0$ and axial anisotropy $0 \le \Delta \le 1$. Total 
spin $S$ is conserved in the isotropic limit, $\Delta = 1$, but only $S_z$ is a good quantum number otherwise. 
The gs is always in the $S_z = 0$ sector for the parameters of interest in this work, and $S_z = 0$ basis states 
are products of $N/2$ spins $\alpha$ and N/2 spins $\beta$. The gs transforms as $P = \pm 1$ 
under the spin parity operator $P$ that reverses all spins and as $C_i = \pm 1$ under inversion 
at sites, which corresponds to reflection through sites $p$, $p+N/2$ in finite systems. 
$H(\alpha,\Delta)$ also has $ C_N$ translational symmetry and inversion symmetry midway between sites. \\

The gs phase diagram of $H(\alpha,\Delta)$ has been studied by 
many-body methods \cite{r1,r14,r16,r17,r18,r19,r20,r14r,r21,r22} such a field theory, 
perturbation theory and density matrix renormalization group (DMRG) that so far 
agree only in part. Broken symmetry is expected and found for ranges of $\alpha$ and $\Delta$, 
with multiple exotic phases near the quantum critical point $\alpha_c = -1/4$, $\Delta = 1$. 
Excitation energies are small, of order $1/N$ for $N$ spins and exceptionally small 
according to one field theory \cite{r23}. One challenge is to distinguish between strictly 
degenerate gs that indicate broken symmetry and nondegenerate gs with tiny excitation energies, 
as discussed carefully by Affleck and Lieb in 1D spin chains \cite{r24}.\\

We adopt a different 
approach that bears directly on broken symmetry and order parameters but only indirectly 
on the phase diagram. We solve $H(\alpha,\Delta)$ exactly for finite $N$ up to $N = 28$. 
Since exact eigenstates respect all symmetries, broken symmetry requires degenerate gs that in 
turn yield order parameters. On the other hand, phase boundaries from finite-size calculations 
 are based on excited-state crossovers, or degeneracy, as discussed for the BOW phase of the 
isotropic AF/AF chain \cite{r3,r25} or of extended Hubbard models \cite{r26,r27} Order parameters 
become nonzero at the boundary and increase in the broken-symmetry phase.  As shown in Section II, 
$H(\alpha,\Delta)$ has doubly degenerate gs at multiple parameter values 
$\gamma_m = (\alpha_m,\Delta_m), m = 1, 2,...,$ where the gs symmetry changes.\\

There are three kinds of gs degeneracy. When $\gamma_m$ corresponds to $C_i = \pm 1$ or $P = \pm 1$, the linear combinations 

\begin{eqnarray}
|\gamma_m \rangle  =(|\gamma_m, 1 \rangle  \pm |\gamma_m,-1  \rangle)/\sqrt{2}    
\label{eq2}
\end{eqnarray}
\noindent have broken inversion or spin parity symmetry and different order parameters. 
When both $C_i$ and $P$ are reversed at $\gamma_m$, either (1,1) and (-1,-1) or (1,-1) and (-1,1), 
the linear combinations in Eq. \ref{eq2} have $C_i P = \pm 1$. There are finite-size 
gaps to all other eigenstates, sometimes remarkably small gaps. Broken $C_i$ symmetry leads as 
usual to a BOW or dimer phase; broken parity P is associated with staggered magnetization 
or N\'{e}el order; broken $C_i$ and P symmetry yields VC order. Dimer, N\'{e}el and 
VC phases are all possibilities for $H(\alpha,\Delta)$ in zero field \cite{r17,r19,r14}. 
Exact diagonalization, albeit limited to discrete $\gamma_m$, makes it possible to compute 
order parameters as well as spin correlations functions or other properties. When systematic 
variations with N are found, extrapolation gives accurate but not exact information about 
the infinite chain. 

Increasing $J_2$ (increasing $-\alpha$) induces a quantum transition from a high 
to low spin. The isotropic chain has a first order transition with discontinuous $S$ 
at $\gamma_1 = (-1/4,1)$. $S_z$ rather than $S$ is conserved in systems with 
axial anisotropy $\Delta < 1$. The normalized spin per site $S(\alpha,\Delta)$ is the gs expectation value
\begin{eqnarray}
2S(\alpha,\Delta)=\frac{2\langle \alpha,\Delta|S^2|\alpha,\Delta \rangle^{1/2}}{[N(N+2)]^{1/2}} \le 1  
\label{eq3}
\end{eqnarray}
\noindent $S(\alpha,\Delta)$ is shown in Section III to be discontinuous for small anisotropy 
$\Delta > \Delta_B \approx 0.95$, continuous for $\Delta < \Delta_B$. 
There is finite N\'{e}el order between $\Delta_A = 0.75$ and $\Delta_B$, 
finite BOW or dimer order for $\Delta < \Delta_A$. The spin transition requires 
exact gs and has apparently not been recognized previously in anisotropic chains. 
Spin correlation function show spiral order at $\gamma_m$ when $C_i$ symmetry is broken, 
as in isotropic AF/AF chains \cite{r25,r28}. We find identical spin correlations functions 
at parameter values $\gamma_m$ when both $C_i$ and $P$ are broken, and interpret these 
results as right and left-handed helices along the unique axis with spins in the xy plane. We 
compare finite-size results with previous theory and find considerable 
agreement as well as occasional disagreement. Our results limit VC degeneracy 
to small ranges of  parameters $\alpha$, $\Delta$  in which $S(\alpha,\Delta)$ varies 
rapidly but continuously. More extended VC phases have been inferred by other methods \cite{r17,r19}.

\section{Degeneracy and broken symmetry}
We summarize some properties of the anisotropic F/AF spin chain, 
Eq. \ref{eq1}, before presenting numerical results for even $N$ with PBC. 
The gs is ferromagnetic at $J_2 = 0$ ($\alpha = 0$) with magnetization in the xy plane 
for $\Delta < 1$. Increasing $J_2 > 0$ (increasing $-\alpha$) induces a quantum 
transition to a low-spin state. The ferromagnetic and singlet gs of the 
isotropic chain are degenerate at $\alpha_c = -1/4$ with energy per 
site $\epsilon_0 = -3/16$. The exact result for the extended system is 
the first degeneracy $\gamma_1 = (-1/4,1)$ for finite $N$, where $S$ 
changes from $N/2$ to 0. Any singlet with $S_z = 0$ can be represented \cite{r29} 
as linear combinations of $N/2$ paired spins $(\alpha_i\beta_j - \beta_i\alpha_j)/\sqrt2 $ 
whose phase is fixed by choosing site $i  < j$. The parity is $P = (-1)^{N/2}$ since reversing 
all spins gives a phase factor of -1 for every singlet pair. The gs linear combination at $\alpha_c$ 
is the uniformly distributed resonating valence bond solid \cite{r15}.\\ 
 
The degeneracy $\gamma_1 = \alpha_1(\Delta)$ between high and low spin shifts to $\alpha_1(\Delta) <\alpha_c = -1/4$. 
The unit step function $2S(\alpha,1)$ at $\alpha_c = -1/4$ decreases rapidly for $\Delta < 1$ and $2S(\alpha,\Delta)$ 
is continuous for $\Delta < 0.95$. In the limit $\Delta = 0$ of extreme anisotropy, the crossover to low spin occurs 
at $\alpha_1(0) = -1/2$ for even $N$. 
The degeneracy $\gamma_1 = (-1/2,0)$ is 
the F/AF version of the Majumdar-Ghosh point of the isotropic AF/AF chain \cite{r4}. Here the exact gs 
is a product of triplets with $S_z = 0$, $(\alpha_i\beta_j +\alpha_i\beta_j)/\sqrt{2}$, either with $i = 2n-1$, 
$j = 2n$ or $i = 2n$, $j = 2n+1$. There are $N/2$ triplets and Eq. \ref{eq3} gives
\begin{eqnarray}
S(-1/2,0)=1/\sqrt{(N+2)}
\label{eq4}
\end{eqnarray}
\noindent The exact result shows that the extended system has $S(\alpha,\Delta) = 0$ for $-\alpha \ge 1/2$ 
over the entire range of $\Delta$. The singlet gs at $\Delta = 1$ becomes a linear combination of 
states with $S^2 \approx N$ in anisotropic chains, and the spin per site goes as $N^{-1/2}$ on the low-spin side.\\

Another relevant limit is $J_1 = 0$, when $H(\alpha,\Delta)$ decouples into chains 
with anisotropic $J_2 > 0$ between nearest neighbors as sketched in Fig. \ref{fig1}. 
$N = 4n$ systems decouple into 1D chains of $2n$ sites that correspond to the XXZ 
Heisenberg spin-1/2 antiferromagnet. The gs of each chain has $S_z = 0$ and there are no 
correlations between spins in different chains. On the other hand, $N = 4n + 2$ systems decouple 
into two radicals with $S_z = \pm 1/2$  whose gs remain entangled even at $J_1 = 0$. We find $4n$, $4n+2$ 
effects in isotropic chains around $\alpha \approx -1/2$, but none at $\gamma_1 = \alpha_1(\Delta)$.\\

We obtain the lowest energy of $H(\alpha,\Delta)$, Eq. \ref{eq1}, for even $N$ and PBC in 
four sectors with $S_z = 0$, $C_i = \pm 1$ and $P = \pm 1$. The absolute gs is degenerate 
\begin{figure}[h]
\begin {center}
\hspace*{-0cm}{\includegraphics[width=7.5cm,height=8.5cm,angle=-90]{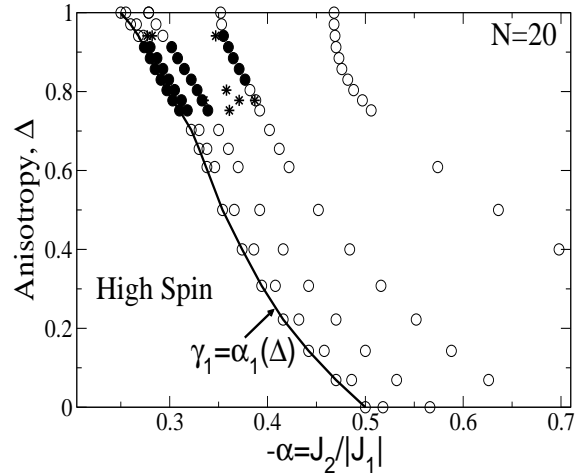}} \\
\caption{Parameter points $(\alpha,\Delta)$ for a doubly degenerate ground state of $H(\alpha,\Delta)$, Eq. \ref{eq1}, with $N = 20$ spins. Open circles indicate inversion degeneracy $C_i = \pm 1$, closed circles indicate spin parity degeneracy $P =\pm 1$, and stars indicate vector chiral degeneracy $C_iP = \pm 1$. The line $\alpha_1(\Delta)$ marks degeneracy between high and low spin.}
\label{fig2}
\end {center}
\end{figure}
\begin{figure}[h]
\begin {center}
\hspace*{-0cm}{\includegraphics[width=7.5cm,height=8.5cm,angle=-90]{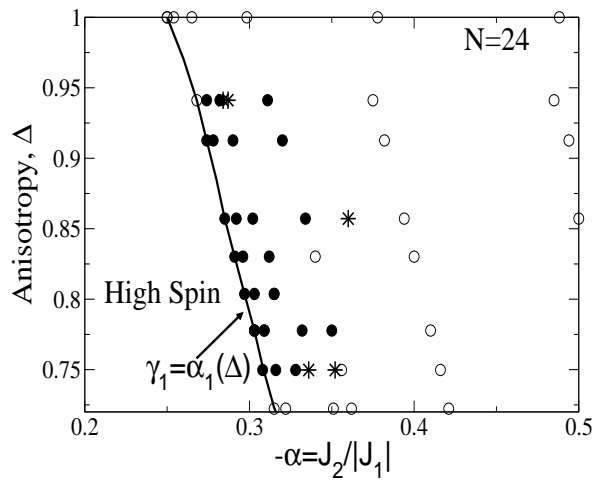}} \\
\caption{Same as Fig. \ref{fig2} for $N = 24$ at intermediate anisotropy $\Delta$ 
and frustration $\alpha$.}
\label{fig3}
\end {center}
\end{figure}
at points $\gamma_m = (\alpha_m,\Delta_m)$ that are found to two or three significant 
figures. Preliminary scans for small $N$ on a rougher grid identifies parameter ranges with 
symmetry crossovers. Fig.  \ref{fig2} shows the $\gamma_m$ in the $-\alpha$, $\Delta$ 
plane for $N$ = 20. Open circles indicate $C_i$ degeneracy, closed circles P degeneracy and 
stars VC degeneracy. The points $\alpha_1$ at $\Delta = 1$ and 0 are exact, independent of 
size, and the line $\gamma_1 = \alpha_1(\Delta)$ hardly depends on $N$. All degeneracy 
for $\Delta < 0.7$ is associated with inversion symmetry. Staggered magnetization or 
VC order is limited to intermediate anisotropy in Fig. \ref{fig2}. The intermediate 
region is expanded in Fig. \ref{fig3} for $N = 24$. \\

Table \ref{tab1} lists the $n-1$ values of $(-\alpha_m,1)$, $m \ge 2$, for isotropic $N = 4n$ 
chains, all with $C_i = \pm 1$. Anisotropic $N = 4n$ chains with $\Delta < 0.7$ also have n 
degenerate points $\alpha_m$ with $C_i = \pm 1$. Broken-symmetry gs are given by Eq. \ref{eq2}. 
When $\gamma_m$ corresponds to $C_i = \pm 1$, the BOW or dimer amplitude $B(\gamma_m)$ is
\begin{table}
\caption {Degenerate ground states at $(-\alpha_m,1)$, $m \ge 2$, of isotropic spin chains, Eq. \ref{eq1}, with $N$ sites and $\Delta = 1$}. 
\begin{tabular}{cccccccc} \hline
$N$ ~& $-\alpha_2$ ~& $-\alpha_3$ ~& $-\alpha_4$ ~& $-\alpha_5$ ~& $-\alpha_6$ &~ $-\alpha_7$ \\\hline
8  ~& 0.342&~      &~      &~      &~      &~       \\ 
12 ~& 0.276&~ 0.401&~      &~      &~      &~       \\
16 ~& 0.260&~ 0.318&~ 0.439&~      &~      &~       \\ 
20 ~& 0.254&~ 0.279&~ 0.352&~ 0.467&~      &~       \\
24 ~& 0.254&~ 0.265&~ 0.299&~ 0.378&~ 0.488&~       \\
28 ~& 0.251&~ 0.259&~ 0.288&~ 0.338&~ 0.381&~ 0.510 \\\hline
\end{tabular}
\label{tab1}
\end{table}

\begin{eqnarray}
B(\gamma_m)=\langle \gamma_m; -1| (S_1 \cdot S_2- S_2 \cdot S_3) |\gamma_m; 1 \rangle/2 
\label{eq5}
\end{eqnarray}
\begin{figure}[h]
\begin {center}
\hspace*{-0cm}{\includegraphics[width=7.5cm,height=8.5cm,angle=-90]{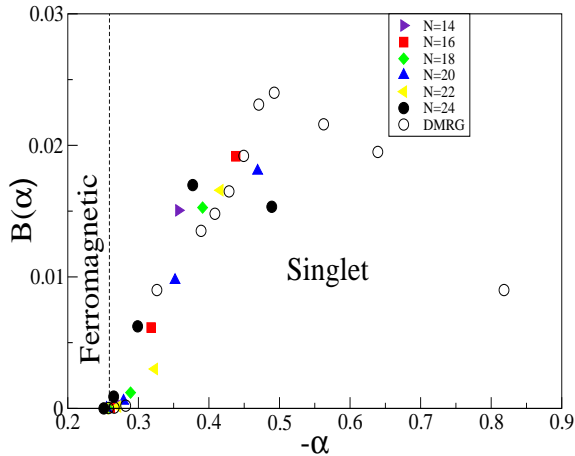}} \\
\caption{(Color online) Discrete order parameter $B(\alpha_m)$, Eq. 5, as a function of frustration $\alpha$ in isotropic $(\Delta = 1)$ chains of $N$ sites; continuous $B(\alpha)$ from DMRG. } 
\label{fig4}
\end {center}
\end{figure}
\noindent The order parameter is the off-diagonal matrix element, $\pm B$ for the two gs, 
and we used translational symmetry. Only BOWs are realized in isotropic $(\Delta = 1)$ systems 
whose $B(\alpha_m,1)$ are shown in Fig. \ref{fig4} for both $N = 4n$ and $4n+2$. $B(\alpha,1)$ has a broad 
maximum around $\alpha \approx -0.5$ and a modest size dependence. DMRG results in Fig. \ref{fig4} 
are based on an algorithm \cite{r30} that adds four spins per step, as needed for accuracy  at $-\alpha > 0.5$, and gives fragments with an even number of sites at each step. The AF/AF chain has a broad maximum \cite{r27} $B(0.4,1) = 0.40$ 
that is 10-fold larger.  The z and transverse parts of $B(\alpha_m)$ are found separately in anisotropic systems with $\Delta < 1$. 
Most $\gamma_m$ in Figs. \ref{fig2} and \ref{fig3} refer to degeneracy under inversion, and $B(-1/2,0) = 1/8$ is exact.\\  

When $\gamma_m$ corresponds to $P = \pm 1$, the amplitude $M_{st}(\gamma_m) > 0$ of the staggered magnetization is

\begin{eqnarray}
M_{st}(\gamma_m)=\frac{1}{2}\langle \gamma_m; -1| (S^z_1 - S^z_2) |\gamma_m;1| \rangle 
\label{eq6}
\end{eqnarray}
\noindent Similarly, when both $C_i$ and $P$ are reversed at $\gamma_m$, the VC order parameters 
$\pm \kappa_z(p;\gamma_m)$ for spins p sites apart are

\begin{eqnarray}
 \kappa_z (p,\gamma_m) &=& \langle \gamma_m; -1,-1| (\vec{S}_1 \times \vec{S}_{1+p})_z|\gamma_m;1,1 \rangle  \nonum \\ 
&=&  \frac{i}{2} \langle \gamma_m;-1,-1|(S^+_1 S^-_{1+p}-S^-_{1} S^+_{1+p})|\gamma_m;1,1 \rangle \nonum \\ 
 & &
\label{eq7}
\end{eqnarray}
The $z$ component of the vector product is finite for axial anisotropy, 
and Eq. \ref{eq7} is for degeneracy between (1,1) and (-1,-1). The matrix element has (-1,1) 
and (1,-1) for the other way of reversing both $C_i$ and $P$. We paid close attention to 
degenerate points $\gamma_m$ and found only two-fold degeneracy. In a few cases, the gap 
to the first excited state is tiny, of the order of $10^{-5}$, far less than $1/N$. Double 
degeneracy implies one broken symmetry in finite chains for any $(\alpha,\Delta)$ in the 
range $-\alpha > 1/4$, $0 \le \Delta \le 1$.  

\begin{figure}[h]
\begin {center}
\hspace*{-0cm}{\includegraphics[width=7.5cm,height=8.5cm,angle=-90]{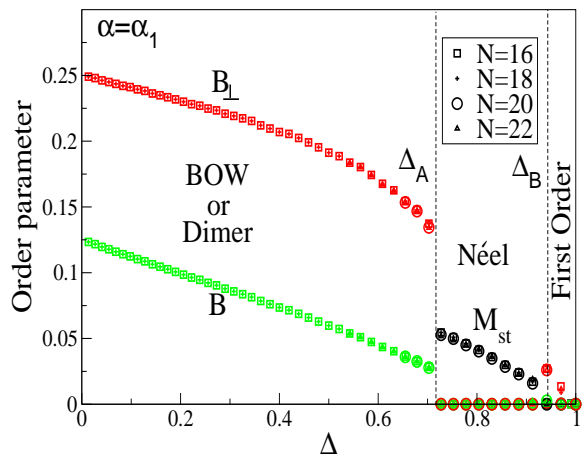}} \\
\caption{(Color online) Order parameters along the degeneracy line 
$\alpha_1(\Delta)$ between high and low spin. $B$ and $B_{\perp}$ are 
Eq. \ref{eq5} and its transverse part, 
and $M_{st}$ is Eq. \ref{eq6}. The spin transition of the infinite 
chain is first order for $\Delta \ge \Delta_B$ and has vanishing 
order parameters.} 
\label{fig5}
\end {center}
\end{figure}

The order parameters $B(\gamma_1)$, its transverse part $B_{\perp}$, and $M_{st}(\gamma_1)$ 
are shown in Fig. \ref{fig5} as a function of anisotropy along $\gamma_1 = \alpha_1(\Delta)$. 
The size dependence is negligible, and $\Delta = 0$ is exact, with $B = 1/8$, $B_{\perp} = 1/4$. 
As mentioned above, finite $B(\gamma_1)$ or $M_{st}(\gamma_1)$ implies that the phase boundary 
where the order parameter becomes nonzero is on the high-spin side of $\gamma_1$, but deviations 
from $\gamma_1$ are only significant for large order parameters. The largest difference occurs 
at $\Delta = 0$, where $\alpha_1 = -0.50$ and  excited-state crossovers give $\alpha = -0.325$. 
The dimer/N\'{e}el boundary in Fig. \ref{fig5} is $\Delta_A(N) = 0.722$ with $\alpha_A = -0.317$ 
for $N$ = 18, 20 and 22, while the N\'{e}el/first order boundary is $\Delta_B(N) = 0.941$ with 
$\alpha_B = -0.2674$. The staggered magnetization increases with anisotropy in the N\'{e}el 
phase to 0.050 at $\Delta_A$. $B(\alpha_1)$ is small $(<0.03)$ 
and decreases with $N$ near 
$\Delta_B$ in the first-order region where we expect $B = 0$ in the extended system. For comparison, 
the infinite time-evolving block decimation (iTEBD) algorithm \cite{r17} gives $\Delta_A = 0.72$, $\alpha_A = -0.320$ and $\Delta_B = 0.93$, $\alpha_B = -0.272$ in Fig. 4 of ref. 17a; the maximum 
N\'{e}el amplitude is 0.055 and $2B_{\perp}$ (called $D^{xy}_{123})$ = 0.35 at $\Delta = 0.65$ where we find 
$2B_{\perp} = 0.309$ at $N =$ 16 and 0.307 at $N =$ 22.  There is remarkably close agreement 
between two completely different calculations. We disagree near $\Delta = 1$ where iTEBD returns 
small finite B rather than B = 0.

\section{Spin transition and correlation functions}

\begin{figure}[h]
\begin {center}
\hspace*{-0cm}{\includegraphics[width=7.5cm,height=8.5cm,angle=-90]{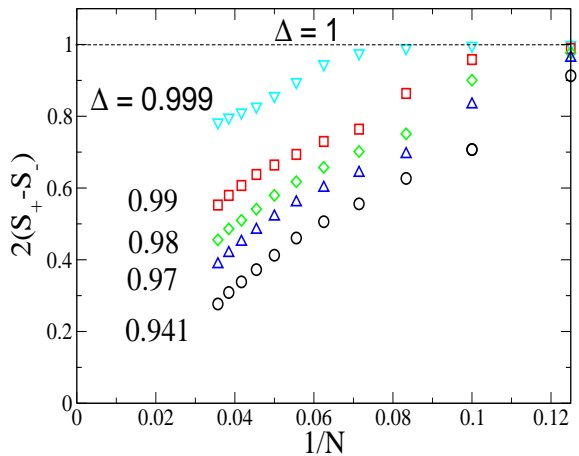}} \\
\caption{(Color online) Discontinuity in $S(\alpha,\Delta)$, Eq. \ref{eq3}, at the 
degeneracy $\alpha_1(\Delta)$ between high and low spin in system of N spins with 
axial anisotropy $\Delta$ up to N = 28.} 
\label{fig6}
\end {center}
\end{figure}

We evaluate the spin per site $2S(\alpha,\Delta)$, Eq. \ref{eq3}, for slightly anisotropic 
systems along the degeneracy $\gamma_1 = \alpha_1(\Delta)$ between high $(S_+)$ and low spin $(S_-)$.
 The size dependence of $S_+(\Delta) - S_-(\Delta)$ is shown in Fig. \ref{fig6} up to $N = 28$. 
The step function of isotropic chains is quickly lost with increasing $N$. Although $1/N$ 
behavior is approximate at best, the gap has vanished by $\Delta = 0.94$ and the extended 
system has a continuous transition for $\Delta < \Delta_B = 0.95 \pm 0.01$. 
The $S(\alpha,\Delta)$ discontinuity at $\alpha_1(\Delta)$ clearly decreases very 
rapidly with anisotropy. The precise value of $\Delta_B$ is less important than 
recognizing a first order quantum transition for $\Delta > \Delta_B$. Orthogonality 
in $S$ then ensures vanishing order parameters for an infinite chain with a first 
order transition in Fig. \ref{fig5}. As an indication of internal consistency, we note 
that almost identical values of $\Delta_B$ are inferred from closing the 
$S_+(\Delta) - S_-(\Delta)$ discontinuity and from the N\'{e}el/first-order boundary.
 
We define gs correlation functions for spins $p$ sites apart using the 
translational symmetry of $H(\alpha,\Delta)$ 

\begin{eqnarray}
C(p,\gamma)&=&\langle \gamma| (\vec{S_1}\cdot \vec{S}_{1+p}) |\gamma| \rangle \nonum \\ 
&=& C_{z}(p,\gamma)_z+C_{\perp}(p,\gamma)
\label{eq8}
\end{eqnarray}

\noindent The $z$ and transverse components are computed separately and are not simply related 
in anisotropic systems. For $S = 1/2$ chains, we have

\begin{eqnarray}
\langle \gamma | S^2_z | \gamma \rangle /N &=&\frac{1}{4}+\sum^{N-1}_{p=1} C_z(p,\gamma)  \nonum \\ 
\langle \gamma | S^2_{\perp} | \gamma \rangle /N & = &\frac{1}{2} + \sum^{N-1}_{p=1} 
C_{\perp}(p,\gamma) 
\label{eq9}
\end{eqnarray}

\noindent Since the gs is in the $S_z = 0$ sector, finite $S(\alpha,\Delta)$ is due to transverse components. 
The high-spin regime with $-\gamma \le \alpha_1(\Delta)$ has $C(p) > 0$ for all $p$, or simply $C(p) = 1/4$ 
at $\Delta = 1$. The low-spin regime necessarily has $C(p) < 0$ for some $p$, and PBC implies an even 
number of sign changes as a function of $p$.  \\

\begin{figure}[h]
\begin {center}
\hspace*{-0cm}{\includegraphics[width=7.5cm,height=8.5cm,angle=-90]{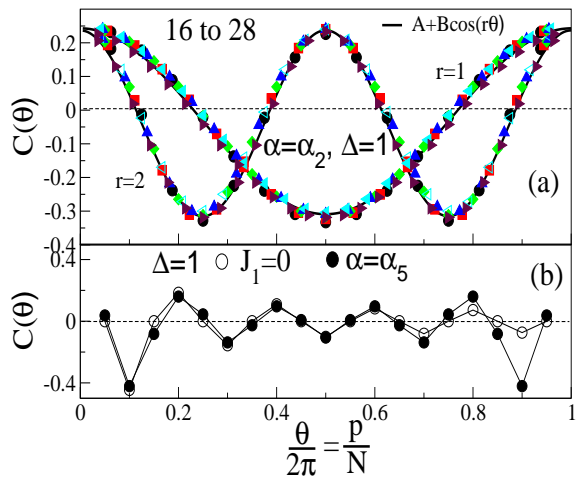}} \\
\caption{(Color online)  Spin correlation functions $C(\theta)$, Eq. \ref{eq8} with $\theta = 2\pi p/N$, 
of isotropic chains $H(\alpha,1)$, Eq. \ref{eq1}. Upper panel: $\alpha_2$ up to 
$N = 28$ with $2r$ changes of sign; the $r = 1$ and 2 lines have $B = 0.276$, $A = -0.0426$. 
Lower panel: closed symbols, $\alpha_5$ for $N = 20$ with 10 sign changes; open symbols, 
decoupled chains with $J_1 = 0$ and $C(p) = 0$ for odd $p$.  } 
\label{fig7}
\end {center}
\end{figure}

The upper panel of Fig. \ref{fig7} shows $C(p,\alpha_2)$ in isotropic chains up to $N = 28$ as 
a function of $\theta = 2\pi p/N$ with $p = 1, 2,...N-1$. At $\Delta = 1$, there are 
two sign changes in the interval $[-1/4,\alpha_2]$ and four sign changes in $[\alpha_2,\alpha_3]$. $C(p,\alpha_2)$ 
is double valued. The systematic size dependence allows inferences about the extended system. 
The line in Fig. \ref{fig7}a has small deviations from sinusoidal due to $C(0) = 3/4$, whose contribution decreases 
as $1/N$. $C(p,\gamma)$ changes sign $2n$ times for $-\gamma > –\gamma_n$, as shown in Fig. \ref{fig7} 
for $N$ = 20, lower panel, for isotropic chains at $\gamma_5 = \alpha_5(1)$. For comparison, we include the 
limit $J_1 = 0$, when $C(p) = 0$ for odd $p$ and has alternating sign for even $p$. The $J_1 = 0$ calculation is for a longer chain that shows decreasing AF correlation with increasing p. Although 
$\alpha_n(1) \approx -1/2$ in Table \ref{tab1} is far from the $J_1 = 0$ limit 
$(\alpha \rightarrow -\infty)$ of decoupled AF chains, 
the spin correlation functions are already similar. They change sign at most $2n$ 
times and account fully for the gs degeneracy of isotropic $N = 4n$ chains. 
Isotropic chains, either F/AF or AF/AF, are limited to BOW or dimer phases, in 
clear disagreement with the VC phase in zero field in Fig. 1a of ref 19 or, 
in a smaller range, in Fig. 2b of ref 17b. \\

\begin{figure}[h]
\begin {center}
\hspace*{-0cm}{\includegraphics[width=7.5cm,height=8.5cm,angle=-90]{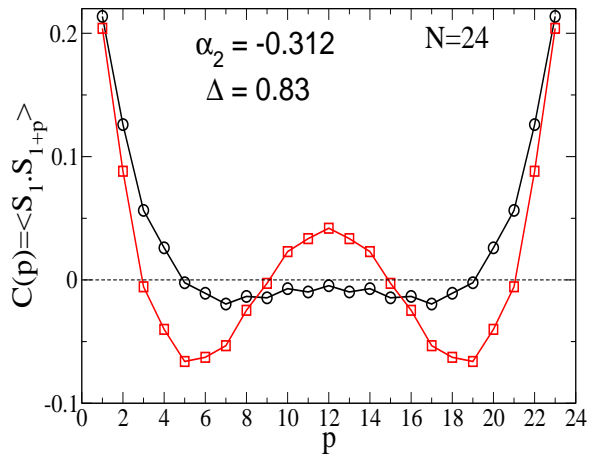}} \\
\caption{(Color online) Spin correlation functions $C(p)$ of a 24-site chain with spin parity degeneracy at 
$\alpha_2 = -0.312$, $\Delta = 0.83$.} 
\label{fig8}
\end {center}
\end{figure}

$C(p;\gamma_m)$ of anisotropic chains are not sinusoidal but still have $2m-2$ and $2m$ 
sign changes when $\gamma_m$ marks broken $C_i$ or $P$ symmetry. The $C(p,\gamma_2)$ in Fig. \ref{fig8} 
are for N\'{e}el order and $N = 24$. There are two and four sign changes as expected. $N = 4n$ 
chains have $n$ points $\gamma_m$ with broken $C_i$ or $P$ symmetry. Additional degeneracy, if any, 
is exclusively due the broken $C_i$ and $P$ symmetry that indicates VC order. Returning to Figs. \ref{fig2} 
and \ref{fig3}, we see that lines $\gamma_m = \alpha_m(\Delta)$ through the mth degeneracy, either $C_i$ or $P$, 
of finite systems partition parameter space into regions in which $C(p)$ changes sign $2m-2$ and $2m$ times. \\

The gs energy per site, $\epsilon_0(\alpha,\Delta)$, is related to spin correlation functions of 
first and second neighbors

\begin{eqnarray}
\epsilon_{0}(\alpha,\Delta)=-C_{\perp}(1)-\Delta C_{z}(1) + \alpha C_{\perp}(2)+ \alpha \Delta C_{z}(2) 
\label{eq10}
\end{eqnarray}
\noindent Degenerate $\epsilon_0(\alpha,\Delta)$ at $\gamma_m = \alpha_m(\Delta)$ is achieved with 
unequal $C(1;\gamma_m), C(2;\gamma_m)$ in Figs. \ref{fig7} and \ref{fig8}. Finite-size 
effects generate small discontinuities in $S(\alpha,\Delta)$ at all $\gamma_m$ with unequal $C(p;\gamma_m)$.  
It is straightforward to count the number of $C_i$ or $P$ degeneracies in finite systems. 
We cannot predict the number of VC degeneracy, the stars in Fig. \ref{fig2} and \ref{fig3},
 and a finer grid may reveal additional gs degeneracy leading to VC order.\\

\begin{figure}[h]
\begin {center}
\hspace*{-0cm}{\includegraphics[width=7.5cm,height=8.5cm,angle=-90]{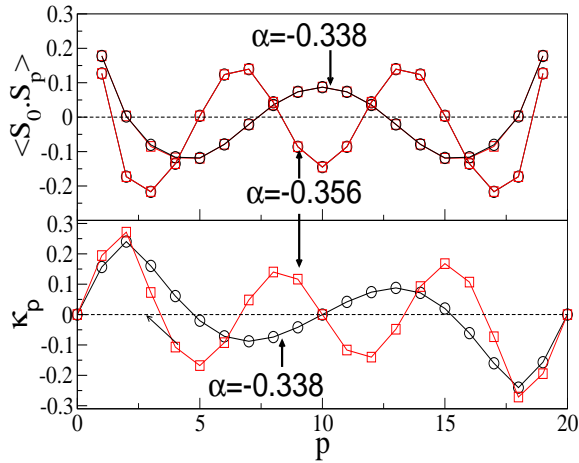}} \\
\caption{(Color online) Spin correlation functions $C(p)$ and order parameters $\kappa_z(p)$, Eq. \ref{eq7}, 
of a 20-site chain with VC degeneracies at $\Delta = 0.75$ and $\alpha = -0.338$ and -0.357. } 
\label{fig9}
\end {center}
\end{figure}

In contrast to P or $C_i$ degeneracy, degenerate $\epsilon_0(\alpha,\Delta)$ under 
reversal of both $C_i$ and $P$ lead to {\it equal} spin correlation functions as shown 
in Fig. \ref{fig9}, upper panel, for $N$ = 20, 
$\Delta = 0.75$, $\alpha = -0.338$ and -0.357. The first point has four sign changes, in accord with 
two $P$ degeneracies at smaller $-\alpha$. There is a $C_i$ degeneracy between the two VC points, 
which accounts for six sign changes at $-\alpha = 0.357$. We always find equal $C(p;\gamma_m)$ at
 VC degeneracies within our 3-4 digit numerical accuracy, except for $N = 4n+2$ systems with $-\alpha > 0.5$ 
where equality is limited to two digits and there are pronounced $4n$, $4n+2$ effects. Equal $C(p;\gamma_m)$ 
for all $p$ implies equal $S(\gamma_m)$ according to Eq. \ref{eq9}. The VC order parameters $\kappa_z(p;\gamma_m)$, 
Eq. \ref{eq6}, in the lower panel of Fig. \ref{fig9} has the same periodicity but they vanish 
at points that are shifted by $\pi/2$. It follows from Eqs. \ref{eq7} and \ref{eq8} that 
when $\gamma_m$ refers to VC degeneracy, 

\begin{eqnarray}
\langle S^+_1S^-_{1+p} \rangle =-C_{\perp}(p,\gamma_m)-i\kappa_z(p,\gamma_m) 
\label{eq11}
\end{eqnarray}
VC order parameters are closely related to transverse spin correlation functions in chains with axial anisotropy.\\ 

A simple classical picture captures the principal features of quantum spins with VC 
order $\pm \kappa_z(p)$. Right and left-handed helices are doubly degeneracy with equal 
energy $\epsilon_0(\gamma_m)$ and spin $S(\gamma_m)$. $H(\alpha,\Delta)$ has axial 
symmetry in spin space and spins in the xy plane for $\Delta < 1$. We picture the two 
helices as $(x_n, \pm y_n) = (cosn \phi, \pm sinn \phi) $ with pitch angle $\phi$ that 
generates a specified number of sign changes between $n = 1$ and $N$. Inversion gives $n \rightarrow  -n$ 
and interchanges the helices. 
Spin parity reverses the z and y components 
of spin, but not 
$s_x = (s^+ + s^-)/2$, and also interchanges the helices. Equal $C(p,\gamma_m)$ follow from projecting 
the helices on a plane that includes the z axis, as illustrated by 
$2\langle x_nx_{n+p}\rangle_n = 2\langle y_ny_{n+p} \rangle_n = cosp\phi$ 
where $<..>_n$ indicates an average over n. The VC order parameters $\kappa_z(p,\gamma_m)$ 
go as $\pm 2\langle  x_ny_{n+p} \rangle_n = \pm sinp \phi $ , shifted by $\pi/2$. \\

When $H(\alpha,\Delta)$ in Eq. \ref{eq1} has classical spins, the gs is easily shown by energy 
minimization to be a spiral with pitch angle $\varphi$ that depends only on $\alpha$,

\begin{eqnarray}
cos\varphi =-J_1/4J_2=-1/4\alpha 
\label{eq12}
\end{eqnarray}
\noindent A finite chain with PBC has $N \varphi = 2\pi m$. The critical point $-4\alpha = 1$ 
corresponds to $\varphi = 0$, 
the ferromagnetic gs. Finite $N = 4n$ limits the remaining angles to $n - 1$ values $0 < \varphi < \pi/2$. 
Spiral spin correlation functions appear automatically in DMRG treatments \cite{r28,r30} of 
isotropic spin-1/2 AF/AF chains. This follows directly from a singlet gs and Eq. \ref{eq8} with 2$C_z(p) = C_{\perp}(p)$, 
although DMRG is not accurate for the sum over all spin correlation functions. A spiral interpretation of 
$C(p,\gamma_m)$ holds in finite F/AF chains at any $\gamma_m$ or, indeed, at any $(\alpha,\Delta)$. 
There are crucial differences, however, between helices and spirals. Left and right-handed helices 
are degenerate at $\gamma_m$ with reversed $C_i$ and $P$ symmetry. Spirals are associated with $C_i = \pm 1$ 
and BOW phases, but are not degenerate in $\pm  \varphi$ in Eq. \ref{eq12}. The pitch angle $\varphi= \phi$ 
of spiral or helices of classical spins follows immediately from PBC. Table \ref{tab1} lists degenerate $\gamma_m$ 
at $\Delta = 1$ for quantum spins that can be converted to angles $\varphi_m$ using Eq. \ref{eq12}. The proper 
number of $\gamma_m$ is found for each $N$, but the actual values differ considerably 
for $-\alpha > 0.4$. Moreover, the points $\gamma_m = \alpha_m(\Delta)$ vary substantially with anisotropy.
 
\begin{figure}[h]
\begin {center}
\hspace*{-0cm}{\includegraphics[width=7.5cm,height=8.5cm,angle=-90]{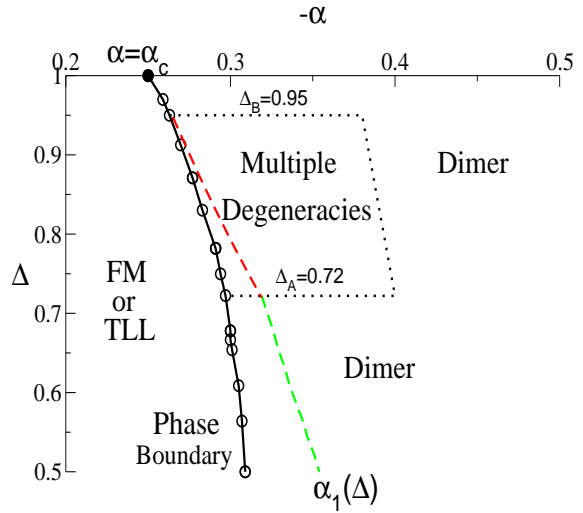}} \\
\caption{(Color online) Approximate gs phase diagram of the anisotropic F/AF chain $H(\alpha, \Delta)$, 
Eq. \ref{eq1}. The phase boundary between high and low spin is an excited-state crossover 
discussed in the text. The degeneracy line $\alpha_1(\Delta)$ has N\'{e}el order between 
$\Delta_A$ and $\Delta_B$, dimer order for $\Delta < \Delta_A$. Dashed lines enclose regions 
with multiple degeneracies (inversion, spin parity, both) where phases are not assigned.} 
\label{fig10}
\end {center}
\end{figure}

\section {Discussion}

Quite unusually for model Hamiltonians, finite spin-1/2 chains with $H(\alpha,\Delta)$ 
in Eq. \ref{eq1}, periodic boundary conditions and frustrated exchanges $J_1 < 0$ and 
$J_2 = \alpha J_1 > 0$ have doubly degenerate ground state (gs) at many parameter values 
$\gamma_m = (\alpha_m, \Delta_m)$. There are three kinds of gs degeneracy: inversion symmetry $C_i = \pm 1$ 
or spin parity $P = \pm 1$ or reversal of both $C_i$ and $P$. We have exploited degeneracy 
for finite $N$ using exact diagonalization to construct broken-symmetry gs according to Eq. \ref{eq2}. 
Exact gs make it possible to compute order parameters for broken symmetry, spin correlation 
functions $C(p,\gamma_m)$ in Eq. \ref{eq8} and the normalized expectation value $S(\alpha,\Delta)$ 
of the total spin in Eq. \ref{eq3}. Access to exact gs properties compensates 
to some extent  for the inherent limitations of finite-size approaches to infinite chains. \\

We have focused on gs degeneracy to construct and characterized broken-symmetry states. 
Quantum phase diagram derived from finite systems invoke excited-state 
degeneracy \cite{r3,r26}. Our finite-size results are summarized in Fig. \ref{fig10}. 
The indicated phase boundary is based on the excited-state degeneracy between first 
excited state in $S_z = 0$ and the $S_z = \pm1$ gs up to $N = 28$, as discussed in related 
1D systems \cite{r2,r26,r27,r28} that conserve $S$. The $N$ dependence of the 
excited-state degeneracy is comparably weak and independent 
of $\Delta$. The order parameter becomes nonzero and an energy gap opens 
very slowly \cite{r3,r27} at a Kosterlitz-Thouless transition. \\

The high-spin phase \cite{r17} is a ferromagnet or a Tomonaga-Luttinger liquid (TLL). 
We find the transition between high and low spin to be first order for $\Delta_B > 0.95$, 
continuous for $\Delta < \Delta_B$. Degeneracy along $\gamma_1 = \alpha_1(\Delta)$ 
is exact at $\Delta = 1$ or 0 and depends very weakly on $N$ in between; as 
seen in Fig. \ref{fig5}, there is N\'{e}el order along along the line 
between $\Delta_A$ and $\Delta_B$, dimer or BOW order for $\Delta < \Delta_A$. 
An almost identical boundary is shown in Fig. \ref{fig4} of ref. 17a or Fig. 2b of ref. 17b, 
based on $\alpha = J_2/|J_1|$ as a perturbation to an exact field theory 
at $\Delta = 0$. As noted in connection with Fig. \ref{fig5}, the order parameters 
of the dimer and N\'{e}el phases also agree well. \\

Multiple degeneracy is limited to intermediate anisotropy $\Delta$ and frustration 
ratio $\alpha $ close to $\alpha_1(\Delta)$. The dashed lines in Fig. \ref{fig10} 
enclose the parameter space in which we find all three kinds of gs degeneracy. 
We cannot assign phases in this region. The degeneracies in Figs. \ref{fig2} and 
\ref{fig3} do not evolve systematically with $N$. Finite-size results  merely place 
  restrictions on phase boundaries. The isotropic ($\Delta = 1$) chain is 
limited to a BOW or dimer phase for $-\alpha< 1/4$, consistent with exclusively $C_i$ 
degeneracy in Table \ref{tab1}. \\

Broken $C_i$ symmetry leading to a dimer or 
BOW phase dominates on the low-spin side. We find $B_z(\gamma_m)$, the $z$ part of the order 
parameter in Eq. \ref{eq5}, to change sign with decreasing $\Delta$, from singlet-type pairing 
at $\Delta = 1$ where the gs is a singlet to triplet-type pairing that is exact at $\Delta = 0$, 
$\alpha = -1/2$. A dimer triplet phase appears for small $\Delta$ in Fig. 2b of ref. 17b, separated 
by a VC phase from a dimer singlet phase. We do not find evidence for VC order outside the dashed lines in 
Fig. \ref{fig10} and consider the sign of $B_z(\gamma_m)$ to be incidental in the dimer phase. The 
small region of multiple degeneracies in Fig. \ref{fig10} is the major difference 
with the extensive VC regions in some \cite{r17,r19,r20,r21} zero-field phase diagrams, 
but not in others \cite{r14,r22}.  \\

Finite-size results complement approximate treatment of extended systems. 
DMRG grows a discrete extended chain while field theory deals with a continuum version of the chain. 
DMRG \cite{r28,r30} results for F/AF or for AF/AF spin-1/2 chains are for open boundary conditions 
and even $N$ that increases by two or four sites per step. Even chains have inversion 
symmetry at the center of the middle bond, but not $C_i$ at any site, and $C_i$ is 
relevant for gs degeneracy. Exact treatment of a half-filled band of $N$ free electrons 
with open boundary conidtions leads to a nondegenerate gs \cite{r27,r30} with a BOW  whose amplitude $B$ 
decreases as $1/N$ and excitation energies of order $1/N$. DMRG works well for model parameters 
leading to substantial $B > 0.01$ such as the broad peak around $\alpha \approx  -0.4$ for 
isotropic F/AF chains in Fig. \ref{fig4}. But DMRG fails in extended 1D systems 
where, as discussed by Affleck and Lieb  \cite{r24}, the distinction 
between gs degeneracy and $1/N$ excitation energies has to be considered.\\
 
Turning to field theory, we note that a continuum model is an approximation, that additional 
approximations are typically needed, and that different field theories are 
possible for $H(\alpha,\Delta)$. Two field theories \cite{r23,r28} for isotropic AF/AF chains 
are not limited to $J_1 > 0$. The anisotropic F/AF chain has been treated with first-order corrections 
in $J_2/J_1$ starting \cite{r17} with exact field theory at $J_2 = 0$ and also as bosonization \cite{r19} in the opposite 
limit of $|J_1| \ll J_2$. The merits of field theory are beyond the scope of this paper. A continuum models 
suppresses the important distinction between inversion symmetry at sites and at the centers of bonds.\\ 
 
We have discussed frustrated chains with axial anisotropy for parameters leading to gs in the $S_z = 0$ 
sector. The energy spectrum of isotropic ($\Delta = 1$) chains in a magnetic field $h =  g\mu_BH$ 
that defines the z axis is simply the zero-field energy plus the Zeeman energy $hm$, 
with $m = \pm 1, \pm 2,~~... \pm S$ for states with spin $S$. The system has axial 
symmetry for $h > 0$ but spin parity is no longer conserved. In strong fields, 
the gs has multipolar phases near the boundary between high and low spin \cite{r19}. 
The general problem of $H(\alpha, \Delta)$ plus a static field is considerably 
more complicated because $S_z$ is conserved only when $h$ is along the unique axis. 
Otherwise, the gs has to be found by separately as a function of $h$.\\ 

In summary, finite F/AF models $H(\alpha,\Delta)$, Eq. \ref{eq1}, with frustration 
ratio $\alpha = J_2/J_1$ and axial anisotropy $0 \le \Delta \le 1$ have doubly 
degenerate gs at multiple points $\gamma_m = (\alpha_m, \Delta_m)$, all in the $S_z = 0$ 
sector. Exact gs make it possible to compute order parameters at $\gamma_m$, 
spin correlation functions and the spin per site, $S(\alpha, \Delta)$. The transition 
from high to low spin with increasing $-\alpha$ is first order for $\Delta  >0.95$, 
continuous for $\Delta < 0.95$, with staggered magnetization or N\'{e}el order 
for $0.72 < \Delta < 0.95 $ and dimer order for stronger anisotropy $\Delta < 0.72$. 
When both $C_i$ and $P$ symmetry are broken, vector chiral (VC) order leads to identical 
spin correlation functions and $S(\alpha,\Delta)$ that is interpreted as right and left-handed 
helices along the unique axis with spins in the xy plane. Finite-size effects are typically 
quite small, small enough to discuss infinite chains. Exact finite-size results are consistent with 
 with many-body treatments aside from limiting a possible VC phase to 
intermediate $\alpha$ and $\Delta$ where $S(\alpha, \Delta)$ varies rapidly but continuously.

{ \bf Acknowledgments:} We thank S. Dutton and 
R. Cava for stimulating discussions of frustrated Cu(II) chains.  
This work was largely performed at the 
TIGRESS high performance computer center at Princeton University 
which is jointly supported by the Princeton 
Institute for Computational Science and Engineering and the 
Princeton University Office of Information Technology. 
We thank the National Science Foundation for partial support 
of this work through the Princeton MRSEC (DMR-0819860).

\end{document}